# Superconducting gap anisotropy in monolayer FeSe thin film

Y. Zhang[1,2], J. J. Lee[1,3], R. G. Moore[1], W. Li[1], M. Yi[1,3], M. Hashimoto[4], D. H. Lu[4], T. P. Devereaux[1,3], D.-H. Lee[5,6] & Z.-X. Shen[1,3,*]

[1]Stanford Institute for Materials and Energy Sciences, SLAC National Accelerator Laboratory, 2575 Sand Hill Road, Menlo Park, California 94025, USA

[2]Advanced Light Source, Lawrence Berkeley National Laboratory, Berkeley, California 94720, USA

[3]Geballe Laboratory for Advanced Materials, Departments of Physics and Applied Physics, Stanford University, Stanford, California 94305, USA

[4]Stanford Synchrotron Radiation Lightsource, SLAC National Accelerator Laboratory, 2575 Sand Hill Road, Menlo Park, California 94025, USA

[5]Department of Physics, University of California at Berkeley, Berkeley, California, 94720, USA

[6]Material Science Division, Lawrence Berkeley National Laboratory, Berkeley, California 94720, USA

* To whom correspondence should be addressed: zxshen@stanford.edu

**Fermi surface topology and pairing symmetry are two pivotal characteristics of a superconductor. Superconductivity in one monolayer (1ML) FeSe thin film has attracted great interest recently due to its intriguing interfacial properties and possibly high superconducting transition temperature ($T_c$) over 77 K[1-12]. Here, we report high-resolution measurements of the Fermi surface and superconducting gaps in 1ML FeSe using angle-resolved photoemission spectroscopy (ARPES). Two ellipse-like electron pockets are clearly resolved overlapping with each other at the Brillouin zone corner. The superconducting gap is nodeless but moderately anisotropic, which put strong constraints on determining the pairing symmetry. The gap maxima locate along the major axis of ellipse, which cannot be explained by a single d-wave, extended s-wave, or $s_{\pm}$ gap function. Four gap minima are observed at the intersection of electron pockets suggesting the existence of either a sign change or orbital-dependent pairing in 1ML FeSe.**

For 1ML FeSe thin film grown on SrTiO$_3$, previous angle-resolved photoemission spectroscopy (ARPES) studies showed that the Fermi surface consists only of electron pockets at the corner of the Brillouin zone and there are no hole pockets at the zone center as typically found in iron-based superconductors[2-4, 10]. Based on such a Fermi surface topology, several pairing symmetries have been proposed theoretically[13-19]: *d*-wave and extended *s*-wave pairing symmetries, where the gaps change sign between different parts of the Fermi surface; s-wave and $s_{\pm}$ pairing symmetries, where there is no sign change on the Fermi surface. In early studies, the superconducting gaps were found to be nearly isotropic on the electron pockets[2, 4], however they provide insufficient information to address the critical issue of pairing symmetry in 1ML FeSe.

Here, we study the 1ML FeSe film using high-resolution ARPES. The film is grown on Nb-doped SrTiO$_3$ and the T$_c$ is around 55K as reported in ref 10. As shown in Fig. 1a, the ARPES intensity map shows one circular-like Fermi surface at M and no Fermi surface at the zone center ($\Gamma$). This result is consistent with previous studies[2-4, 10]. However, by performing high-resolution measurements at the M point using particular photon polarizations, two ellipse-like electron pockets are clearly resolved (Figs. 1c and 1d). According to the band calculations, the Fermi surface consists of one ellipse electron pocket at each Brillouin zone boundary in one-iron Brillouin zone[20, 21] (Fig. 1b). When the glide-mirror symmetry of the iron-selenium plane is considered, the unit cell doubles and the Brillouin zone folds[22]. As a result, the horizontal ellipse pocket ($\delta_1$) folds onto the vertical ellipse pocket ($\delta_2$) in the two-iron Brillouin zone (Fig. 1b). This is consistent with the observed Fermi surface topology at the M point (Figs. 1c and 1d). The photoemission intensity of the main band $\delta_2$, is much higher than that of the folded band $\delta_1$ at the bottom corner of the two-iron Brillouin zone.

Clear delineation of Fermi surface is important for determining the gap structure in 1ML FeSe. Figure 2 shows the superconducting gaps associated with the $\delta_1$ and $\delta_2$ electron bands along the $\Gamma$-M direction. Two electron bands are clearly resolved in the second derivative images along the M-$\Gamma$ direction (Fig. 2a). The orbital selectivity of different polarizations enables us to probe different bands separately[22, 23], which is crucial for measuring the superconducting gaps in a multi-band system. By using linear vertical polarization (LV), the $\delta_1$ and $\delta_2$ bands can be probed selectively by choosing the cut momenta at either the left or bottom corner of the Brillouin zone (Fig. 2b). The back bendings of bands are clearly observed attesting for the high data quality. The gap magnitude is determined by the gap fitting of symmetrized EDCs at the relevant $k_F$'s (Figs. 2c and 2d), *i.e.*, the gap minima of the corresponding band dispersions (Fig. 2a). The superconducting gap is ~10 meV for the $\delta_1$ band and ~13 meV for the $\delta_2$ band.

The multi-gap behavior on $\delta_1$ and $\delta_2$ suggests that the superconducting gap is anisotropic. By choosing different experimental setup, we can selectively measure the superconducting gap on different sections of the electron pockets (Fig. 3). The gap anisotropy is obvious near the major axis of the ellipses (Figs. 3a ~ 3f), and manifested by the maxima along the 90° and 270° directions in Figs. 3c and 3f, respectively.

Because the Fermi surface of 1ML FeSe is C4 symmetric, we could then map the gap measurement data onto one ellipse electron pocket. The results are shown in Figs. 4a ~ 4b. The smooth evolution of the gap tied to the underlying Fermi surface is consistent with it being a superconducting gap. Because ARPES measures the absolute value of the gap function we perform a best fit of the observed energy gap as a function of angle around the Fermi pocket by $|f(\theta)|$. The result of such fitting gives $f(\theta) = (9.98 \pm 0.10) - (1.24 \pm 0.13) \cos 2\theta + (1.15 \pm 0.13) \cos 4\theta$ meV.

One characteristic of the gap anisotropy is that the gap maxima locate along the major-axis of the ellipse (the 90° and 270° directions). We simulated the gap distribution on the $\delta_2$ electron pocket using several trigonometric gap functions that are proposed under d-wave, $s_\pm$ and extended s-wave pairing symmetries (Figs. 4d ~ 4f). For the $|(\cos k_x - \cos k_y)/2|$ and $|\cos k_x \cos k_y|$ gap functions, the pair strength is strongest at the center of the ellipse electron pockets meaning that the gap maxima should locate along the minor-axis of the ellipse (0° and 180° directions) (Figs. 4e and 4f). This contradicts to our observation. For the $|(\cos k_x + \cos k_y)/2|$ gap function, it generates gap maxima along the correct directions. However, the $|(\cos k_x + \cos k_y)/2|$ gap function is not energy favored in 1ML FeSe because the pairing strength is strongest at the center and corner of the one-iron Brillouin zone where there is no Fermi surface. In order to achieve the 8~13 meV gap magnitude on the electron pockets, large and unrealistic gap value (> 100 meV) needs to be used in the simulation. Therefore, the observed momentum locations of the gap maxima cannot be explained by a single trigonometric gap function. Mixing of different gap functions with the same symmetry[18] or the high harmonic gap functions should be considered[14]. Alternatively, the gap anisotropy on the $\delta_2$ ellipse electron pocket is C2 symmetric, which does not violate the crystal symmetry. Therefore, the observed gap distribution can be consistent with an s-wave gap function if the pairing strength varies on the Fermi surface following the crystal symmetry of material.

Another characteristic of the gap anisotropy is the presence of four gap minima along the 45° directions of the ellipse. On one hand, the gap minima can be induced by the Fermi surface hybridization near the intersection of the $\delta_1$ and $\delta_2$ ellipse electron pockets. According to the band calculations, either the spin-orbital coupling or the breaking of glide-mirror symmetry would lift the band degeneracy at the Fermi surface crossing of two ellipses in two-iron Brillouin zone[25, 26]. For 1ML FeSe film grown on SrTiO$_3$, finite spin-orbital coupling is present as observed in bulk FeSe[27] and the glide-mirror symmetry no longer exists due to the presence of substrate. Therefore, it is expected that there is finite hybridization between the $\delta_1$ and $\delta_2$ ellipse electron pockets. If gap changes sign near the intersection of $\delta_1$ and $\delta_2$, such as the gap under d-wave and extended s-wave pairing symmetries, the hybridization between $\delta_1$ and $\delta_2$ would mix the gaps with opposite sign and generate gap minima or gap nodes depending on the strength of the hybridization[13, 15]. On the other hand, the gap minima may originate from the multi-orbital nature of superconductivity. The electron pockets are constructed by $d_{xz}/d_{yz}$ and $d_{xy}$ orbitals and the orbital character varies when moving around the Fermi surface[20-23]. The gap minima locate along the 45° directions of the ellipses where the mixing of $d_{xz}/d_{yz}$ and $d_{xy}$ orbitals is strongest. Gap would be minima if the intra-orbital pairing dominates the pairing interaction in 1ML FeSe[18].

**Method**

FeSe films were grown on high quality Nb-doped (0.05% wt) SrTiO$_3$ (100) substrates. TiO$_2$ terminated atomic flat surface were prepared by degassing at 450 °C for several hours and subsequently annealing at 900 °C for 20 min. Ultrahigh purity selenium (99.999%) was evaporated from an effusion cell with a thermal cracking insert (Createc) while iron (99.995%) was evaporated from a 2 mm rod using an electron beam evaporator (Specs). The growth was carried out under Se-rich condition with a Se/Fe flux ratio of 3~4. Substrate temperatures were kept at 380 °C during the growth. The films were subsequently annealed at 450 °C for four hours immediately after growth. The films were transported to the Stanford Synchrotron Radiation Lighsource Beamline 5-4 via a vacuum suitcase with base pressure of $5 \times 10^{-10}$ Torr.

ARPES measurements were performed at the Stanford Synchrotron Radiation Lightsource Beamline 5-4. The data were taken with 22 eV photons. The temperature was kept at 20 K for the superconducting gap measurement. The overall energy resolution was 5 meV, and the angular resolution was less than 0.3 degree. All the samples were measured in ultrahigh vacuum with a base pressure better than $3 \times 10^{-11}$ Torr.

**References**

1. Wang, Q.-Y. *et al.* Interface-Induced High-Temperature Superconductivity in Single Unit-Cell FeSe Films on SrTiO$_3$. *Chin. Phys. Lett.* **29**, 37402-037402 (2012).

2. Liu, D. *et al.* Electronic origin of high-temperature superconductivity in single-layer FeSe superconductor. *Nat. Commun.* **3**, 931 (2012).

3. He, S. L. *et al.* Phase diagram and high temperature superconductivity at 65 K in tuning carrier concentration of single-layer FeSe films. *Nat. Mater.* **12**, 605–610 (2013).

4. Tan, S. *et al.* Interface-induced superconductivity and strain-dependent spin density waves in FeSe/SrTiO$_3$ thin films. *Nat. Mater.* **12**, 634-640, (2013).

5. Zhang, W.-H. *et al.* Direct Observation of High-Temperature Superconductivity in One-Unit-Cell FeSe Films. *Chin. Phys. Lett.* **31**, 017401 (2014).

6. Peng, R. *et al.* Tuning the band structure and superconductivity in single-layer FeSe by interface engineering. *Nat. Commun.* **5**, 6044 (2014).

7. Deng, L. Z. *et al.* Meissner and mesoscopic superconducting states in 1-4 unit-cell FeSe films. *Phys. Rev. B* **90**, 214513 (2014).

8. Ge, J.-F. *et al.* Superconductivity in single-layer films of FeSe with a transition temperature above 100 K. *arXiv*, 1406.3435 (2014).

9. Xiang, Y.-Y., Wang, F., Wang, D., Wang, Q.-H. & Lee, D.-H. High-temperature superconductivity at the FeSe/SrTiO$_3$ interface. *Phys. Rev. B* **86**, 134508 (2012).

10. Lee, J. J. *et al.* Interfacial mode coupling as the origin of the enhancement of $T_c$ in FeSe films on SrTiO$_3$. *Nature* **515**, 245-248, (2014).

11. Coh, S., Cohen, M. L. & Louie, S. G. Structural template increases electron-phonon interaction in an FeSe monolayer. *arXiv*, 1407.5657 (2014).

12. Timur, B. & Marvin, L. C. Effects of charge doping and constrained magnetization on the electronic structure of an FeSe monolayer. *Journal of Physics: Condensed Matter* **25**, 105506 (2013).

13. Hirschfeld, P. J., Korshunov, M. M. & Mazin, I. I. Gap symmetry and structure of Fe-based superconductors. *Rep. Prog. Phys.* **74**, 124508 (2011).

14. Maier, T. A., Graser, S., Hirschfeld, P. J. & Scalapino, D. J. d-wave pairing from spin fluctuations in the K$_x$Fe$_{2-y}$Se$_2$ superconductors. *Phys. Rev. B* **83**, 100515 (2011).

15. Mazin, I. I. Symmetry analysis of possible superconducting states in $K_xFe_{2-y}Se_2$ superconductors. *Phys. Rev. B* **84**, 024529 (2011).

16. Hao, N. & Hu, J. Odd parity pairing and nodeless antiphase $s^{\pm}$ in iron-based superconductors. *Phys. Rev. B* **89**, 045144 (2014).

17. Yang, F., Wang, F. & Lee, D.-H. Fermiology, orbital order, orbital fluctuations, and Cooper pairing in iron-based superconductors. *Phys. Rev. B* **88**, 100504 (2013).

18. Saito, T., Onari, S. & Kontani, H. Emergence of fully gapped $s_{++}$-wave and nodal d-wave states mediated by orbital and spin fluctuations in a ten-orbital model of $KFe_2Se_2$. *Phys. Rev. B* **83**, 140512 (2011).

19. Lin, C.-H., Chou, C.-P., Yin, W.-G. & Ku, W. Orbital-Parity Distinct Superconducting Pairing Structures of Fe-based Superconductors under Glide Symmetry. *arXiv*, 1403.3687 (2014).

20. Kuroki, K. *et al.* Unconventional Pairing Originating from the Disconnected Fermi Surfaces of Superconducting $LaFeAsO_{1-x}F_x$. *Phys. Rev. Lett.* **101**, 087004 (2008).

21. Graser, S., Maier, T. A., Hirschfeld, P. J. & Scalapino, D. J. Near-degeneracy of several pairing channels in multiorbital models for the Fe pnictides. *New J. Phys.* **11**, 025016 (2009).

22. Brouet, V. *et al.* Impact of the two Fe unit cell on the electronic structure measured by ARPES in iron pnictides. *Phys. Rev. B* **86**, 075123 (2012).

23. Zhang, Y. *et al.* Orbital characters of bands in the iron-based superconductor $BaFe_{1.85}Co_{0.15}As_2$. *Phys. Rev. B* **83**, 054510 (2011).

24. Norman, M. R., Randeria, M., Ding, H. & Campuzano, J. C. Phenomenology of the low-energy spectral function in high-$T_c$ superconductors. *Phys. Rev. B* **57**, R11093-R11096 (1998).

25. Borisenko, S. et al. Direct observation of spin-orbit coupling in iron-based superconductor. *arXiv*, 1409.8669 (2014).

26. Lin, C.-H. et al. One-Fe versus Two-Fe Brillouin Zone of Fe-Based Superconductors: Creation of the Electron Pockets by Translational Symmetry Breaking. *Phys. Rev. Lett.* **107**, 257001 (2011).

27. Zhang, P. et al. Observation of two distinct $d_{xz}/d_{yz}$ band splittings in FeSe. *Phys. Rev. B* **91**, 214503 (2015).

**Acknowledgements**

ARPES experiments were performed at the Stanford Synchrotron Radiation Lightsource, which is operated by the Office of Basic Energy Sciences, U.S. Department of Energy. The Stanford work is supported by the US DOE, Office of Basic Energy Science, Division of Materials Science and Engineering. D.-H. L. was supported by the U.S. Department of Energy, Office of Science, Basic Energy Sciences, Materials Sciences and Engineering Division, grant DE-AC02-05CH11231.

**Author Contributions**

Y. Z. and Z. X. S conceived the project. Y. Z., J. J. L., R. G. M., W. L. and M. Y. took the ARPES measurements. Y. Z. analyzed the data. W. L., J. J. L. and R. G. M. grew the films. M. H. and D. H. L. assisted in ARPES measurements at Stanford Synchrotron Radiation Lightsource. T. P. D. D. H. L. and Z.X. S supervised the project. Y. Z., D. H. L. and Z. X. S wrote the paper with input from all coauthors.

**Author Information**

The authors declare no competing financial interests. Correspondence and request for materials should be addressed to Z.-X. Shen (zxshen@stanford.edu).

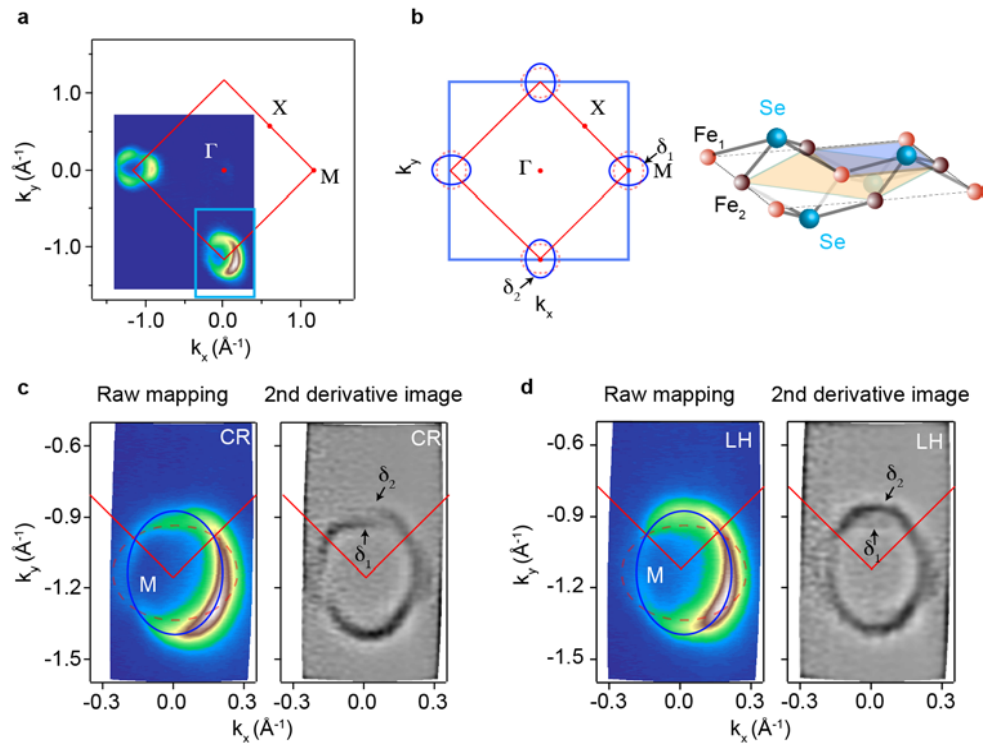

**Figure 1 | Fermi surface topology. a**, Fermi surface mapping of 1ML FeSe taken in circular (CR) polarization. The Brillouin zone is defined by the two-iron unit cell with the Γ-M direction along the iron-iron bond direction. **b**, The one- and two- iron unit cells in the iron-pnictogen/chalcogen plane (right panel) and the corresponding one- and two- iron Brillouin zones and Fermi surfaces (left panel). **c**, Fermi surface mapping (left panel) and its second derivative image (right panel) taken in CR polarization at ~120 K using high energy and momentum resolution. The derivation was taken along both $k_x$ and $k_y$ direction. **d** is the same as **c**, but taken with linear horizontal (LH) polarization. The in-plane polarization vector is along the $k_x$ direction.

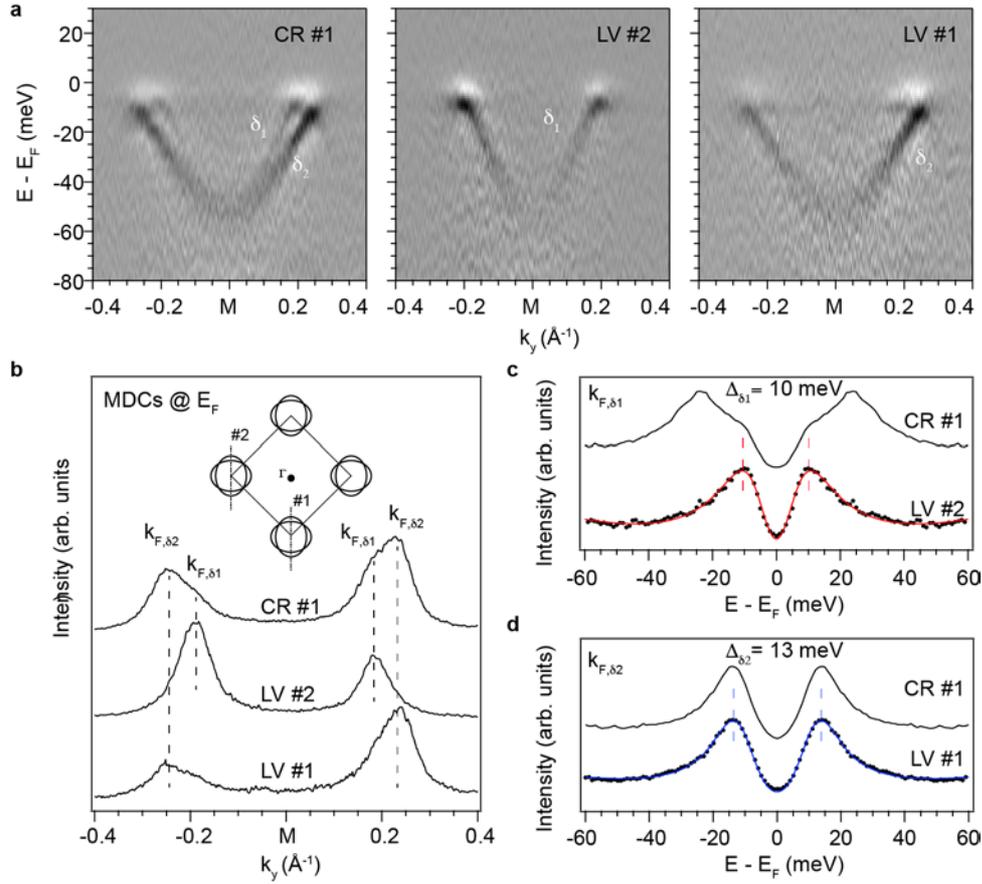

**Figure 2 | Multi-gap behavior of the superconducting state. a**, The second derivative images of photoemission spectra taken in CR and linear vertical (LV) polarization at 20K. The in-plane polarization vector is along the $k_y$ direction for LV polarization. The momenta of cut #1 and cut #2 are shown in the inset of **b**. **b**, The corresponding MDCs taken at the Fermi energy ($E_F$) of the data in **a**. **c**, Symmetrized EDCs taken at the Fermi crossings ($k_F$'s) of the $\delta_1$ electron band with CR and LV polarization. The gap magnitude was obtained by fitting the symmetrized EDCs (black dots) using a phenomenological superconducting spectra function[24]. The fitting result is shown by the red solid line. **d** is the same as **c** but taken at the $k_F$'s of the $\delta_2$ electron band.

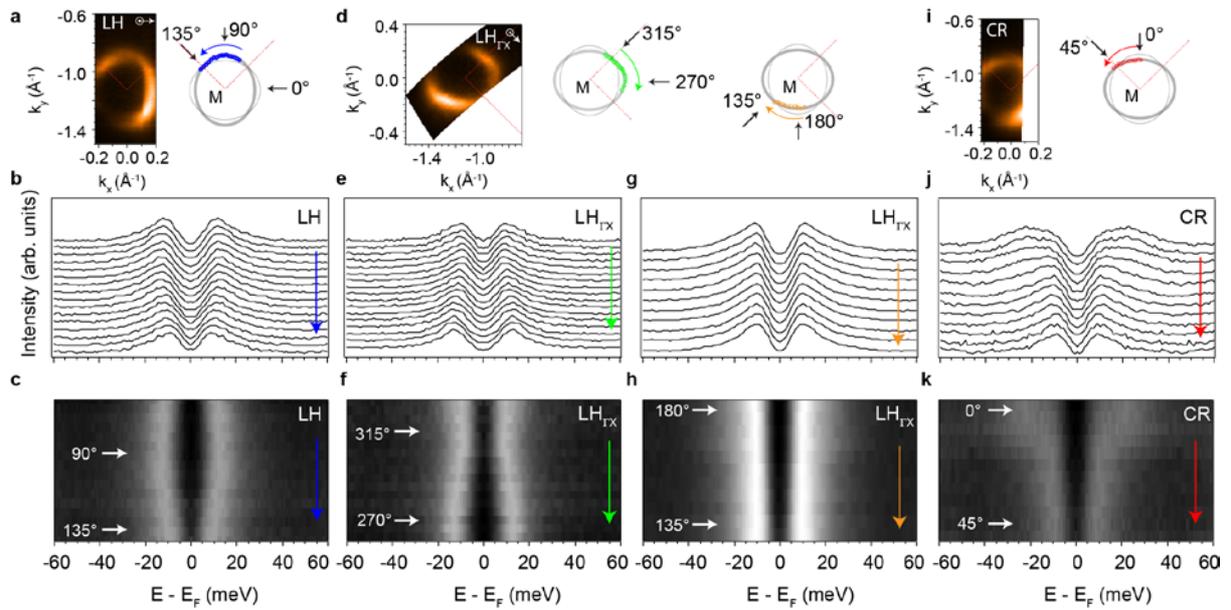

**Figure 3 | Momentum dependence of the superconducting gaps on the electron pockets.**

**a**, The Fermi surface mapping taken in LH polarization. The polarization vector is shown in the inset. **b**, Symmetrized EDCs taken on one section of the electron pockets. The momenta of the data points are shown in **a**. **c**, Merged image of the symmetrized EDCs in **a**, which highlights the variation of the superconducting peak positions. **d** – **k** are the same as **a** and **c**, but taken on different sections of the electron pockets.

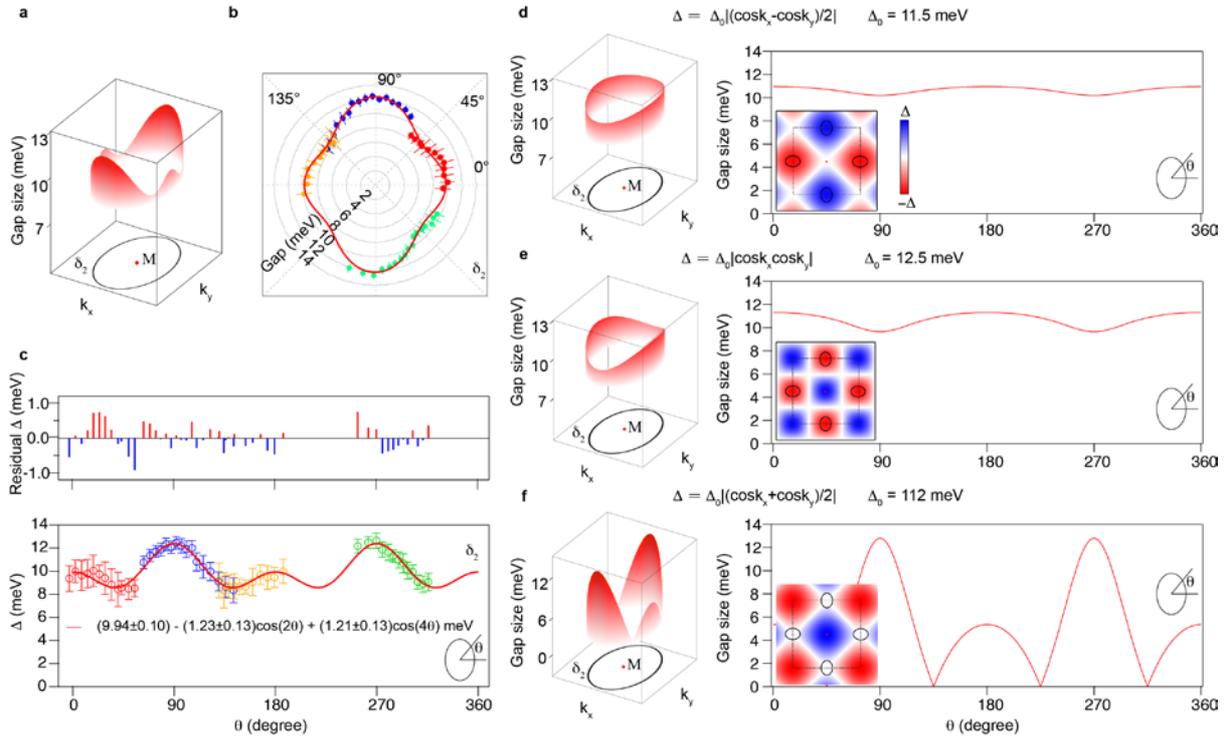

**Figure 4 | Superconducting gap anisotropy in 1ML FeSe. a** and **b**, Superconducting gap anisotropy on the ellipse-like electron pocket $\delta_2$. The gap magnitudes were obtained by fitting the symmetrized EDCs in Fig.3 using a phenomenological superconducting spectral function[24]. The gap measurement data were mapped onto $\delta_2$ ellipse pocket according to the C4 rotational symmetry. The error bars are estimated by the fitting process. **c**, Phenomenology function fitting of the gap anisotropy in 1ML FeSe. The residual values are shown in the upper panel presenting the fitting quality. The fitting result is shown by the red solid line with the function $(9.98 \pm 0.10) - (1.24 \pm 0.13)\cos 2\theta + (1.15 \pm 0.13)\cos 4\theta$ meV. **d**, Simulation of the gap anisotropy on the $\delta_2$ electron pocket using the $|(\cos k_x - \cos k_y)/2|$ gap function. The inset panel illustrates the sign change on the Fermi surface. **e** and **f** are the same as **d** but calculated using $|\cos k_x \cos k_y|$ and $|(\cos k_x + \cos k_y)/2|$ gap functions.